\documentclass[]{llncs}
\usepackage{mathrsfs}
\usepackage{subfigure}
\usepackage{graphics}
\usepackage{balance}
\usepackage{graphicx}
\usepackage{pifont}

\usepackage{amsfonts}

\usepackage{pifont}

\usepackage{epstopdf}
\DeclareMathSizes{10}{9}{6}{4}

\usepackage{etoolbox}
\makeatletter
\patchcmd{\maketitle}{\@copyrightspace}{}{}{}
\makeatother
\begin{document}

\title{Gaming the Game: Honeypot Venues Against Cheaters in Location-based Social Networks}

% Title portion
\author
{ 
Konstantinos Pelechrinis, Prashant Krishnamurthy, Ke Zhang
\\
School of Information Sciences \\
University of Pittsburgh\\
{\em \{kpele, prashant, kez11\}@sis.pitt.edu}
}
\institute{}

\maketitle
\pagenumbering{arabic}
 
\begin{abstract}
The proliferation of location-based social networks (LBSNs) has provided the community with an abundant source of information that can be exploited and used in many different ways.   
LBSNs offer a number of conveniences to its participants, such as - but not limited to - a list of places in the vicinity of a user, recommendations for an area never explored before provided by other peers, tracking of friends, monetary rewards in the form of special deals from the venues visited as well as a cheap way of advertisement for the latter.  
However, service convenience and security have followed disjoint paths in LBSNs and users can misuse the offered features.  
The major threat for the service providers is that of fake check-ins.  
Users can easily manipulate the localization module of the underlying application and declare their presence in a counterfeit location.  
The incentives for these behaviors can be both earning monetary as well as virtual rewards.  
Therefore, while fake check-ins driven from the former motive can cause monetary losses, those aiming in virtual rewards are also harmful.  
%despite the actual motivations of the cheating users, fake check-ins can (i) cause monetary losses to participating venues and (ii) degradations of 
In particular, they can significantly degrade the services offered from the LBSN providers (such as recommendations) or third parties using these data (e.g., urban planners).    
In this paper, we propose and analyze a honeypot venue-based solution, enhanced with a challenge-response scheme, that flags users who are generating fake spatial information.  
We believe that our work will stimulate further research on this important topic and will provide new directions with regards to possible solutions. 
\end{abstract}

%\vspace{1mm}
%\noindent
%{\bf Categories and Subject Descriptors:} 
%C.2 {[Computer - Communication Networks]}, 
%C.2.0 {[General]}: {Security and Protection}, 
%C.2.4 {[Distributed Systems]}: {Distributed Applications}
%
%\vspace{1mm}
%\noindent
{\bf General Terms/Keywords:} 
Location-based Social Networks, Fake check-ins, Honeypots
%
%\vspace{1mm}
%\noindent
%{\bf Keywords:} 
%%\keywords{
%Proportional Fair scheduler, Misreporting attack 
\section{Introduction}
\label{sec:intro}

During the last couple of years, a new class of digital social networks, namely, location-based social networks (LBSNs), have enjoyed rapid proliferation\footnote{While we are mainly focusing and referring to location-based social networks in this paper, our work is applicable to any location-based service regardless of the presence of an explicit social network or not.  However, the majority of these services integrate a strong social component.}.  
These communities not only have interests in common, but they are also bounded with regards to their geographic location (e.g., same city).  
Despite their recent deployment, these services have enjoyed large adoption.  
For instance, Foursquare just one year after being launched had been valued by venture capitals at \$100 million \cite{fsq-value}!  
Google is also upgrading it's own LBSN (Google Latitude) enhancing it with more features, while Facebook recently acquired Gowalla showing its interest to create a competitive upgrade of its own location component, namely, Facebook Places. 
Yelp as well has recently added a location component to each mobile application, through which users can check-in to places, keeping up with the rest location-based services.   
LBSNs offer a number of convenient features.  
For example, every user can easily get information with regards to his geographic location on demand. 
In addition, each user can track his friends, which can potentially help him to explore new places.  
Recently, LBSNs launched recommendation engines that aggregate all the information from all the users' check-ins.  
Moreover, as a recent case study has shown \cite{lindqvist11}, the gaming aspects of these systems (e.g., earning points for visiting places) form an important motivation for people to adopt their usage.  

Furthermore, the geographical data that are generated for the users of these mobile applications, can facilitate studies that span a huge spectrum of fields.  
Human mobility, urban planning, epidimiology and spatial business planning are just a few examples.  

Nevertheless, this wide utilization of LBSNs is accompanied with security threats from possible misuse of the systems.  
While there have been many user concerns (e.g., \cite{girls-around-me}) and research (e.g., \cite{smokescreen} \cite{wosn10}) related to privacy issues, in this work we are considering another major risk in LBSNs, that of {\bf fake check-ins}.  
While it is only recently that it has received the required attention from the research community, it is a legitimate concern and source of frustration for both the users (e.g., \cite{user-furstrations}) as well as the providers (e.g., \cite{fsq-concerns}).  

\subsection{Cheating Types and Their Effects}

Similar to the different incentives of users to share their locations, there are also different reasons why a user, say Bob, will report fraudulent whereabouts.  
There are two major cheating types that we will briefly describe in the following; (i) monetary cheaters and (ii) gamer cheaters.  

Some location-based services, offer Groupon-like deals to its users \cite{fsq-amex}; discounts are offered for check-ins to specific venues participating in such campaigns.   
He {\em et al.} \cite{he11} have reported that the majority of the special offers (more than 90\%) in Foursquare - the major LBSN to date - require multiple check-ins (e.g., $X$ times) to the venue.  
In other words, if Bob wants to {\em unlock} this deal, he needs to visit the locale $X-1$ times before being able to liquidize it.  
Hence, he is tempted to generate a number of fake check-ins in order to obtain the offer faster and with less cost.  
We will call such cheaters, {\bf monetary cheaters} in the rest of this paper.  
Clearly, monetary cheaters can lead to revenue losses for the establishements that offer this deal as we have shown in our recent work \cite{lbsn12}.

Furthermore, almost all LBSNs have integrated gamification techniques in order to keep the attention of their users.  
For instance, in Foursquare a user is able to earn points for every check-in, badges for a specific series of check-ins, ``mayorships'' of venues when he has the most check-ins in the venue within the last two months etc. \cite{fsq-points}. 
Google Latitude has included a leaderboard as well \cite{latitude-points}.   
A large fraction of users view these virtual rewards as means to {\em prove} their social status (e.g., more mayorships translate to a more outgoing, social person etc.) \cite{fsq-gamer-cheater} and hence form an important reason for them to continue using the system.  
Nevertheless, this forms an additional incentive for Bob to fake his presence, that is, to check-in to  places to simply earn more virtual rewards.   
We will refer to such cheaters as {\bf gamer cheaters}.  
While, with a first thought gamer cheaters do not seem to cause any damage to the system, this is not actually correct.  
In particular, LBSN providers make use of the aggregated check-in information across all users to provide services such as location recommendations to their users \cite{fsq-recommendation}.  
The existence of many fake check-ins (regardless if they are from monetary or gamer cheaters), will add noise in the input of the underlying recommendation engines, and hence the offered service will be significantly degraded.  
Furthermore, while it is not clear what is the exact effect of ``noisy'' (i.e., fake) data on the interdisciplinary studies mentioned above, it should be evident that erroneous information will lead to results that do not represented the actual reality.

Our data crawled from Foursquare, indicate that only 162,147 of the 27,219,001 venues existing at the time of crawling, i.e., {\bf 0.6\%} of the total venues, offer special deals to their customers.  
Combined with the fact that the majority of the LBSNs do not currently support Groupon-like offers, renders the gamer cheaters the biggest misbehaving threat and the central focus of our work.

Monetary and gamer cheating check-ins can be further classified in two categories with regards to the distance between the pretended to be in venue and the actual location of the cheater (say Bob).  
In {\em far away} fake check-ins, Bob is actually located much further (e.g., more than 2 miles) from the locale he checks in, while in {\em near by} fake check-ins, he is located fairly close to, but not in, the venue he declares.  
The notion of near and far is relevant to the accuracy of the positioning technology used from the application.    
In other, words for far away check-ins the distance between Bob and the fake location is much larger than the localization error, while in near by fake check-ins this distance is within the localization error.  
 
In theory, the former type can be easily caught, by simply verifying that the GPS coordinates reported from the mobile device correspond to that of the corresponding venue.  
Actually, Foursquare has developed the {\em cheater code} \cite{cheater-code} in an effort to minimize fake spatial information, which among other operations, performs this sanity check.   
The cheater code imposes additional rules on users' check-in frequency and speed. 
Nevertheless, the open nature of the operating systems of the mobile devices, have emerged a number of applications that alter the GPS coordinates that are passed to the corresponding application (e.g., ``Fake Location'' for iPhone \cite{fake-location}, ``Fake GPS location'' from Android \cite{fake-gps-location} etc.).  
Such applications, render the above detection schemes not efficient.  
Since every fake check-in can be successfully pretended as taking place near the corresponding venue, only the near by check-ins are relevant for our study. 

As it might be evident, fake check-in detection is crucial for the long-run success of the LBSN paradigm.  
While location-proofs that have been proposed in other contexts can be applied for verifying a check-in, there are various reasons - which we will analyze in the following section - that render them a not so attractive solution within the context of these mobile applications.  
In this paper, we propose a novel approach for detecting location cheaters in LBSNs.  
Our proposed solution is mainly based on the primitives of honeypots.  
In brief, the location-based service provider can create a number of well-targeted (not existing)  honeypot venues (HV), which are attractive to the cheaters (e.g., easy to obtain mayorship).  
Honest users are not expected to check-in to these HVs and hence, whoever check-ins is flagged as a potential cheater.  
HVs are extremely efficient with gamer cheaters.  
However, they might not perform as well with monetary cheaters as explained later.     
Even though the latter represent an extremely small percentage of misbehaving users, we enhance our initial HV scheme with a challenge-response mechanism tailored to monetary cheaters.

\begin{figure*}[t]
\begin{center}
\vspace{-1.5in}
\includegraphics[scale=0.4,angle=90]{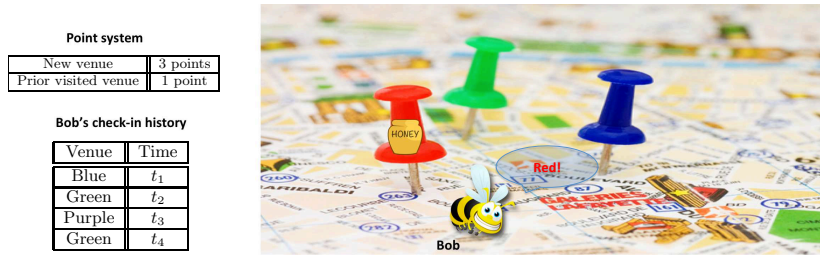} \vspace{-1in}
 \caption{Pictorial representation of the honeypot venues.}\label{fig:hv}
 %\vspace{-0.3in}
\end{center}
\end{figure*} 

%Note here that, we are mainly focused on gamer cheaters.  
%Almost all LBSNs have deployed gamification techniques as compared to a lesser penetration rate of special offers in these systems.  
%Nevertheless, we will enhance the HV design with a challenge-response scheme appropriate for monetary cheaters.  

{\bf Scope of our work: }
Gamification of the World Wide Web and mobile applications is being increasingly used to engage users \cite{gamification}.  
For instance, Q\&A social networks (e.g., Yahoo! Answers, Quora.com etc.) form their own game, which of course can be manipulated by users who simply want to ``win'' the game.   
Hence, gaming cheaters will be present and relevant in a variety of settings. 
However, this will increase the amount of non useful information present in these systems and invevitably reduce their credibility and appeal.  
Our framework, while presented in this paper with a focus on location-based services, it can form the basis for cheating detection schemes in other domains with appropriate modifications.

The rest of the paper is organized as follows.  
Section \ref{sec:related} describes work related with location verification.  
In Section \ref{sec:system} we present and analyze our cheating detection scheme, while Section \ref{sec:conclusions} concludes our work.

%mention that locations proofs have been traditionally utilized but requires many times the presence of a third party, not accurate, or in the case of the location proofs requires capable hardware etc. --> all the work is done from the provider of the service ;) 
% categorize two kinds of cheaters: gaming and monetary.  In each one of them you have nearby and far.  Far can be caught from location cheating in principle - however faking GPS apps can overcome this as well.  Location proofs can be in place (say examples) however this has problems above.  A novel approach (at least for the gaming cheaters - which are the biggest of all(?)) can be honeypot venues.
% challenge response in the sense: what is the special today?
% say that the two type of cheaters have different behaviors
% analysis in the sence of threshold for checkins at HV, probability of checkin an HV for better desing of them, what do you do after you flag them (e.g., let them checkin but not use them for recomendation or not even increase counters)
\section{Related studies}
\label{sec:related}

In this section we briefly discuss related with our work studies.  In particular, we survey literature on location proofs and secure localization.  

{\bf Location proofs: }
With the increased importance of spatial information for various applications, location-proofs have gained attention in the research community during the last years. 
Saroiu and Wolman \cite{Saroiu2009} define location proofs as {``a piece of data that certifies a device to a geographical location''}.  
While there are many ways proposed in the literature to generate these meta-data, in general location proofs are based on cryptographic primitives and are issued by trusted infrastructure (e.g., WiFi APs in combination with a trusted third party).

%
%Recently, several applications and systems have been proposed to verify the location claims. 
%Foursquare develops the $cheater$ $code$\cite{cheater-code} to impose additional rules on users' check-in frequency and speed in consideration of the low-accuracy and vulnerability of GPS-based location verification mechanism, thus mitigating potential opportunities of fake check-ins to some degree. However, cheaters can easily bypass this code, such as modifying GPS API and automating a cheating process while clearly identifying these rules.
%
For instance, Denning and MacDoran \cite{Denning1996} describe a location-based authentication system where the position at any time is uniquely identified by a location signature.  
The signature is created by a location signature sensor (LSS) and it is time varying, hence, making it difficult to be forged.   %since the signature changes with time. 
However, this system relies on a dedicated hardware and requires auxiliary equipment to strengthen the weak GPS signal in indoor environment.
Saroiu and Wolman \cite{Saroiu2009} design a scheme where location proofs are handed out by WiFi access points. Each mobile device signs the APs' beacons and send them back to APs. The latter upon reception of the signed beacon  creates a location signature for the mobile user. %While, this scheme does not need to rely on GPS and extra devices but a tough work to update APs' current software and preconfigure APs' physical coordinates. 
Zhang {\em et al.} \cite{Zhang_Li_Trappe_2007} aslo utilize WiFi infrastructure and design a power modulated challenge-response location verification system. This mechanism utilizes RF signal strength from multiple APs to verify whether the claimed location is within the overlapping range of neighboring APs. %The mechanism also needs to prefigure APs especially their related locations to special locales. Thus, it is unrealistic to be deployed in LBSNs with a large-scale environment.
%Furthermore, Kj\ae rgaard and Wirz \cite{Kjrgaard2012} present a clustering approach to detect indoor flocks of mobile users (i.e., spatio-temporal clusters). 
Luo and Hengartner \cite{luo10}, after presenting six essential goals that a location proof system should follow, propose a scheme, which is based on cryptographic hashes and WiFi APs.  
The proposed system, is able to also retain the user's privacy.  

While the above schemes have not been designed with LBSNs in mind, they can be tweaked in order to be applicable in this context.  
Recently, we have designed a prototype fake check-in detection scheme, based on the primitives of location proofs \cite{lbsn12}.  
In brief, we utilize the notion of location signature using WiFi infrastructure enhancing it with the notion of flocks for identifying users that are not at the location and time they claim to be at. 
%
%In the context of LBSN, as aforementioned He {\em et al} \cite{he11} have identified the problem of fake check-ins, without providing any solution to it.  
%Foursquare has developed the {\em cheater code} \cite{cheater-code} in an effort to minimize fake spatial information.  
%The cheater code imposes additional rules on users' check-in frequency and speed.  % in consideration of the low-accuracy and vulnerability of GPS-based location verification mechanism, thus 
%However, this mitigates potential fake check-ins to some extent only, since cheaters can easily bypass this detection \cite{he11}.  %code, such as modifying GPS API and automating a cheating process while clearly identifying these rules. 
%Our preliminary system design, is based on the primitives of location-proofs and can be complementary to efforts such as the cheater code.  
%In brief, we utilize the notion of location signature using WiFi infrastructure enhancing it with the notion of flocks for identifying users that are not at the location and time they claim to be at.  
%This work focuses on clustering a location proof set into different spacial subsets in order to efficiently detect the appearance of flocks. In our work, we also utilize spacial clustering method but mainly aims to classify fake or true location claims (check-ins) by provided proofs. 
%
However, there are some important limitations on the integration of location proof-based solutions with LBSNs.  
In particular, such schemes rely on the existence of third-party (trusted) infrastructure that can distribute and/or verify the location proofs.  
The cost for deploying this infrastructure is not trivial.  
Furthermore, location proofs are often based on the received signal strength at the mobile device.  
However, different hardware have different capabilities with respect to the accuracy of these  measurements and this can significantly affect the performance of the system.  
Finally, the fact that wireless signals are not geo-fenced, make it extremely hard to distinguished  location proofs issued to a mobile device very close to, but not in, the venue claimed.  
Hence, we are interested in a detection scheme that (i) does not rely on third-party infrastructure and (ii) its performance is not tight to the hardware of the mobile device and the distance between the latter and the venue.   

{\bf Secure localization: } 
Secure localization has been extensively studied within the context of wireless sensor networks.  
Nevertheless, there are studies that apply the same principles to tag user-generated content with spatial timestamps that verify their location.  
For example, Lenders {\em et al.} \cite{lenders08} propose a combination of secure localization with a certification service in order to assign a level of trust on the geo-tagged user content, while preserving privacy.  
Capkun and Hubaux \cite{capkun06} propose {\em verifiable multilateration}, which is based on distance bounding.  
In brief, the position of a device is infered based on its distance from a set of known location reference points (at least three).  
Other approaches belonging to this category (e.g., \cite{denning96} \cite{otason05} \cite{haeberlen04}) exhibit drawbacks similar to that of location-proofs.  
In particular, requirements such as dedicated infrastructure (e.g., reference points) or reliance on the cooperation with telecom providers (e.g., localization through cellular base station) makes these schemes prohibitive for deployment with location-based services at the immediate future. 

In the following section we propose a novel fake check-in detection scheme, borrowing ideas from the traditional computer security field - and in particular the fundamendals of honeypots. 
The proposed system satisfies both of the above design requirements.

\section{Our Proposed Scheme}
\label{sec:system}

In this section, we will present our proposed system for identifying possible cheating users.  
We would like to emphasize on the fact that this scheme is mainly designed as a filtering mechanism, flagging users with high or low levels of suspicious behavior.  
To reiterate, our main focus is gamer cheaters.  
However, we adjust the proposed HV scheme for dealing with monetary cheaters as well.  
%Due to the different incentives between monetary and gamer cheating users, their behavior also differs, and hence separate mechanism should be designed for each class of misbehaviors.  

{\bf Our scheme in a nutshell: }
In brief, gamer cheaters are attracted by venues that can facilitate their goal for as many as possible virtual rewards.  
In other words, they do not care for the specifics of the venues as long as the latter satisfy their goals.  
Hence, the LBSN service provider can create {\em fake} venues - the {\bf honeypots} - that appear attractive to gamer cheaters (e.g., for the case of Foursquare a possible honeypot venue is one that appears to be easy to obtain its ``mayorship'').  
Given that under honest use of the system no one should be present in that locale, users that check-in to honeypot venues are automatically flagged as (potential) gamer cheaters.  

On the contrary, monetary cheaters are clearly attracted by venues that offer special deals.  
While, the service provider could create fake venues with special offers, this might not be attract the majority monetary cheaters. 
The latter are interested in real world rewards and thus, they are most probably focused to venues they are already aware of.  
Therefore, on top of the HV mechanism, we propose to integrate a challenge-response scheme during the check-in process.  
This can verify with high probability the actual presence of the person in the locale.  
Challenges will be isseued only to venues with special offers, to keep the overhead on the check-in process minimal.

\subsection{Honeypot Venues VS Gamer Cheaters}

In computer security, honeypot refers to a machine that appears to be part of the local network and exhibits known and obvious vulnerabilities, which an attacker could exploit to penetrate in the network.  
Nevertheless, this machine is in reality isolated from the rest of the network and monitored from the network administrator.  
Honeypot machines are deployed to detect any malicious attempts to hack into the network and to obtain passive information for the practises of the attackers \cite{honeypots}.

We are utilizing the same idea for {\em trapping} (gamer) cheating LBSN users.  
Figure \ref{fig:hv} visually presents our system.  
Bob, interested in obtaining as many points as possible, obtains a list of venues from the system.  
In this specific example, Bob is awarded with 1 point for every check-in to an already ``visited'' venue, and 3 points for every check-in to a new place.    
Since the blue and the green venues appear in Bob's check-in history, they offer less points for checking-in as compared to the red honeypot venue, which is new to him\footnote{Of course the details of the virtual rewards are different for different systems, but HVs can always be created with the principle of being more appealing to gamer cheaters as compared to real locales.}.  

In general the design of the HV should be such that maximizes the check-in probability $P_{HV}$ of a cheater at venue HV.  
A simple {\bf behavioral model} for this probability could be the following.
Let us consider $C$ to be the set of appearling features for a gamer cheater.  
For simplicity let us consider two representative features drawn from the Foursquare paradigm; the number of possible points $n$ earned from the check-in and the probability $m$ of becoming the ``mayor'' of the venue, i.e., $C=\{n,m\}$.  
Then the probability $p$ should be thought as a function of the elements of $C$:

\begin{equation}
P_{HV} = f(C)=f(n,m)
\label{eq:prob}
\end{equation}  
 
\begin{figure}[t]
\begin{center}
\vspace{-0.6in}
\includegraphics[width=6cm,angle=90]{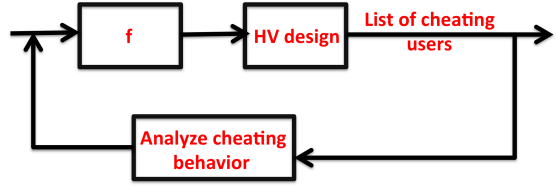} \vspace{-0.7in}
 \caption{Reiforcement learning of model $f$.}\label{fig:behavioral}
\vspace{-0.3in}
\end{center}
\end{figure}

In the case considered, the function $f$ should clearly be a non-decreasing function with respect to both variables $n$ and $m$.  
Of course, the exact shape of $f$ is not known, but it can be reinforced by observing the behavior of identified cheaters.  
This constant feedback process, will facilitate the design of more effective HVs and is graphically depicted in Figure \ref{fig:behavioral}.  
As we can see the behavioral model $f$, and as a consequence the deployed HVs, is regularly refined through analyzing the behavioral data of already identified cheaters. 

Let us further explain the above through a simple example and assume that our initial model is linear.   The initial model $f$ further assigns equal importance to characteristics $n$ and $m$ for the decision of a cheater to perform a check-in.  
In other words, assuming $k$ total possible venues to check-in, the probability that a cheater checks-in to venue $i$ is given by:

\begin{equation}
p_i=\frac{\alpha\cdot n_i + \beta \cdot m_i}{\sum_{j=1}^k \alpha\cdot n_j + \beta \cdot m_j} 
\label{eq:model_ex}
\end{equation}     

\noindent where $\alpha=\beta=0.5$. 
Therefore, the overall probability $p_c$ that a cheater checks-in a HV is given by: 

\begin{equation}
p_c = \sum_{i\in H} p_i
\label{eq:overall}
\end{equation}  	

\noindent where $H$ is the set of HVs. 

After designing the HVs based on this model, we will eventually identify a number of cheaters, say, Bob, Eve and Jack.  
By analyzing their check-in history, we can refine Equation \ref{eq:model_ex}.  
For instance, the data might show that ``mayorships'' are more important and hence it should be $\beta > \alpha$, or that cheaters prefer checking-in to expensive restaurants (e.g., in order to be associated with a ``higher social status''). 
The latter would change the set $C$ by adding as a new element the type $g$ of the HV ($C=\{n,m,g\}$).

An extreme case arises when $C=\emptyset$.  
This means that gamer cheaters do not pay attention to specific features of the venues, but they just blindly check-in to places at random.  
%In this case, in order to maximize the probability of a gamer cheater checking-in to a HV, we need to have a large number of honeypots.  
%In particular, 
In this case, if there are $\lambda$ HVs and $\phi$ real venues, we get:

\begin{equation}
p_c=\frac{\lambda}{\lambda+\phi}
\label{eq:1}
\end{equation}    

%Hence, in order to maximize $P_{HV}$ we need to have $\lambda \gg \phi$.  
%A same requirement can arise, 
The same holds true, when all venues around the misbehaving user are equally appealing, i.e., $p_i = p, \forall i\in\{1,...,k\}$.  

Of course, even honest users can accidentally check-in to a HV.  
However, these instances are not expected to be excessive.  
Therefore, a {\bf suspiciousness level} $l(u)$, can be defined for every user $u$.  
$l(u)$ can be a function $h$ of many factors such as the number of checkins to HVs $q(u)$ of user $u$ and the number of distinct HVs $r(u)$ that $u$ has checked-in at\footnote{One can define more factors that contribute to the calculations of $l(u)$.}, that is, $l(u)=h(q(u),r(u))$.   
By defining a threshold $L$, we can have the following decision rule for gamer cheaters:

\begin{equation}
l(u) > L \Rightarrow u~is~flagged~as~cheater
\label{eq:decision}
\end{equation}

Identifying the optimal value for $L$ is of course not trivial.  
A low value for $L$ can lead to many false positives (i.e., honest users flagged as cheaters), while a large value for $L$ can increase the detection time and/or increase the false negatives (i.e., misbehaving users considered honest).  

{\bf Advanced response from cheaters: }
It should be evident, that if dishonest users become aware of this HV system, they can be more cautious with their check-ins.  
Their goal should be to keep their suspiciousness level below the threshold $L$.  
However, note here that, even if our scheme might not be able to flag them as cheaters (e.g., $l(u)<L$), it will have significantly contained the number of fake check-ins, since the only way for $u$ to keep his suspiciousness level low is by reducing his fake activity.  
This is true, given the fact that honeypot venues look exactly like legitimate locales and hence, the ability of a cheater to identify HVs is at best random.  

It is clear from the above design and analysis of the HV scheme, that the proposed system exhibits the following attractive characteristics:

\begin{itemize}
\item It does not require the cooperation of third parties (e.g., certification providers, telecom providers etc.) and can be deployed and controled purely by the LBSN provider.
\item It is neither hardware dependent, not does it require special hardware at the mobile devices of the end users.  Hence, it is ideal for immediate deployment.  
\end{itemize}

\subsection{Challenge-Response for Monetary Cheaters}

As aforementioned, monetary cheaters are mainly interested into venues with special offers.  
This means that we should include in the set $C$ one or more variables with regards to the special deals of the HV.  
For instance, the number of special offers $x$ and their type ${\bf y}$ (i.e., requiring multiple check-ins, majority of check-ins etc.) are two staightforward candidates\footnote{Vector ${\bf y}$ is a $1 \times s$ binary vector, where $s$ is the number of different possible types of offers.  Element $y_i$ is 1 iff there is an offer of type $i$.}.  
Consequently, honeypot venues specifically tailored to this kind of cheaters should be designed by assigning a much higher weight to the corresponding variables (e.g., $x$ and ${\bf y}$).  

However, monetary cheaters are by default interested in real-world rewards.  
Hence, they are expected to only consider establishments that they have visited in real life or locales they are aware of and they would be interested in visiting.  
In order to increase the efficiency of the HV scheme against this class of dishonest users, we enhance   it with a challenge-response mechanism.  

\begin{figure}[t]
\begin{center}
\includegraphics[width=6cm,angle=90]{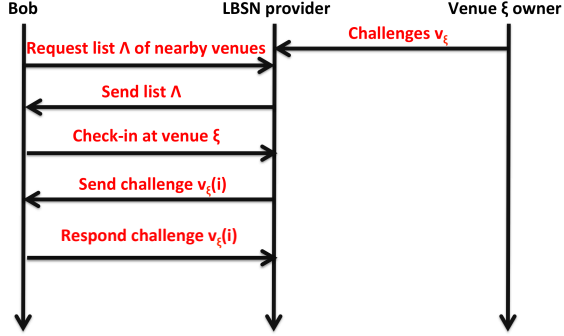}%\vspace{-0.3in}
 \caption{The challenge-response scheme for monetary cheaters.}\label{fig:challenge}
% \vspace{-0.3in}
\end{center}
\end{figure}

In particular, every time Bob, attempts to check-in a venue $\xi$ which offers a special deal, a question relevant to the venue is asked, and he will have to pick the answer from a given menu.  
The question is randomly selected from a set of challenges ${\bf v}_{\xi}$. 
Of course, this challenge should be relevant to both the venue as well as the time of check-in.  
For instance, the application could ask for the ``special of the day'' for the case of a restaurant.  
As it is evident, the load of issuing the challenges should be distributed to the locale owners.  
The latter should decide how often they change these challenges based on the tradeoff between monetary loss and cost of seting up the detection scheme.  
Figure \ref{fig:challenge} depicts the challenge-response process.

While, relying on the venue owner can still leave an open door for fake check-ins (e.g., locale owners do not change the challenges often, or they do not set them up at all), recall that monetary fake check-ins represent a small only fraction of the cheating behavior.  
Hence, the fake information present in the system is expected to be small compared to its initial level after the deployment of both HVs and challenge-response.  
This significantly icrease the value of the data, since they are truthfull and can be reliable used for services and studies as aforementioned.  

On a more philosophical note, monetary cheating exhibits indirectly a preference signal of misbehaving  user for specific venues.  
The thesis behind this claim is that he would not be engaged into cheating if he did not have even the slightest preference towards the specific locale.  
%   (either through being interested in trying out a new establishment or via returning somewhere he has been in the past in real life).  
Therefore, the degradation caused from monetary cheating in services such as recommendations might be less as compared to the case of gaming fake check-ins\footnote{Of course, the thurthfulness of this statement heavily depends on the specific algorithms used for recommendations.}.

\vspace{1mm}
{\large {\bf Evaluations}}

We would like to emphasize on the fact that we consciously did not perfom evaluations of the proposed schemes.  
Thorough evaluations would require the creation of a large number of honeypot venues.  
%In particular the overall probability $p_c$ that a cheater checks-in a HV is given by:
%
%\begin{equation}
%p_c = \sum_{k\in H} p_i
%\label{eq:overall}
%\end{equation}  	
%
%\noindent where $H$ is the set of HVs.  
Equations (\ref{eq:overall}) and (\ref{eq:1}) give the overall probability of a cheater checking-in to a HV.  
Of course, the ultimate probability of flagging a user as cheater depends on the function $h$ and the threshold $L$.  
However, it is evident that in order to maximize the probability that a cheater check-ins to a HV, a large number of honeypot venues, comparable to the number of real venues, might be required ($\lambda \gg \phi$).  
For that, we did not want to interfere with the operations of the LBSNs providers, who are able to have perfect control over HVs, if they were to deploy them.  

In the near future we plan to collaborate with an LBSN provider in an effort to examine the viability of the proposed scheme.

\vspace{1mm}
{\large {\bf Discussion}}

While our work deals with the detection of cheating users with regards to the generated check-ins, it is also important for the location-based service provider to decide what measures it should take against them.  
It is not necessary that these users are banned from the system.  
However, the latter can simply ignore the data generated from them in any service that provides and requires input from user generated data.  
Additionally, it can flag them as ``fake'', for notifying third parties that might utilize them to be cautious.  
In general, the policy followed can differ for different systems and depends on the provider.  

However, a more subtle issue arises if we consider the way that the system should {\em promote} HVs to users.  
For example, if $\delta$ venues are presented to the user who wants to check-in how many of those should be honeypots? 
In what order are they presented? 
On the top of the list? 
Towards the bottom? 
Uniformly interleaved?
The answers to these questions might be customized to every user $u$ and be a function of $l(u)$.  
Providing answers to these questions is outside the scope of our work, but it is clearly an issue that needs to be carefully thought.  
  
\section{Conclusions}
\label{sec:conclusions}

Applications that facilitate location information to provide a number of novel services, have emerged during the last years.  
However, these applications have mainly focused on providing users with an easy way to generate huge volumes of data, mainly through the action of check-in.  
This has left the floor open for misbehaving users to game the system and even overwhelm it with fake geographical information. 
Filtering non truthful spatial information is hence, of critical importance. 
In this paper, we propose a novel scheme for detecting fake check-ins in location-based services.  
Our system is based on the primitives of honeypots.  % (and challenge-response).  
As compared to other possible solutions (e.g., location proofs and secure localization) it possesses the advantage that it can be solely deployed by the location-based service provider without the need for - trusted - third party entities.
%
%\begin{tabular}{|c||c|}
%\hline Venue & Time \\ 
%\hline
%\hline Blue &  $t_1$ \\
%\hline Green & $t_2$ \\ 
%\hline Purple & $t_3$ \\
%\hline Green & $t_4$  \\
%\hline 
%\end{tabular} 
%
%\begin{tabular}{|c||c|}
%\hline New venue &  3 points \\
%\hline Prior visited venue & 1 point \\
%\hline 
%\end{tabular} 

\balance
\vspace*{1.5mm}
\small
\bibliographystyle{unsrt}
%\balance
\bibliography{main}

%\appendix
%\input{ap}

\end{document}